\documentclass[aps,prl,twocolumn,showpacs]{revtex4}
\usepackage{graphicx}
\usepackage{times}
\usepackage{latexsym}
\def\beq{\begin{eqnarray}}
\def\eeq{\end{eqnarray}}
\def\bey{\begin{eqnarray}}
\def\eey{\end{eqnarray}}

\def\lsim{\mathrel{\raise.3ex\hbox{$<$\kern-.75em\lower1ex\hbox{$\sim$}}}}
\def\gsim{\mathrel{\raise.3ex\hbox{$>$\kern-.75em\lower1ex\hbox{$\sim$}}}}

\newcommand{\be}{\begin{equation}}
\newcommand{\ee}{\end{equation}}

\begin{document}

\title{The PAMELA and ATIC Signals From Kaluza-Klein Dark Matter}  
\author{Dan Hooper$^{1,2}$ and Kathryn M. Zurek$^{1,3}$}
\address{$^1$Particle Astrophysics Center, Fermi National Accelerator Laboratory, Batavia, IL 60510 \\
$^2$Department of Astronomy and Astrophysics, The University of Chicago, Chicago, IL 60637\\
$^3$Department of Physics, University of Michigan, Ann Arbor, MI 48109 }

\date{\today}

\begin{abstract}

In this letter, we study the possibility that Kaluza-Klein dark matter in a model with one universal extra dimension is responsible for the recent observations of the PAMELA and ATIC experiments. In this model, the dark matter particles annihilate largely to charged leptons, which enables them to produce a spectrum of cosmic ray electrons and positrons consistent with the PAMELA and ATIC measurements. To normalize to the observed signal, however, large boost factors ($\sim$\,$10^3$) are required. Despite these large boost factors and significant annihilation to hadronic modes (35\%), we find that the constraints from cosmic ray antiproton measurements can be satisfied. Relic abundance considerations in this model force us to consider a rather specific range of masses (approximately 600-900 GeV) which is very similar to the range required to generate the ATIC spectral feature. The results presented here can also be used as a benchmark for model-independent constraints on dark matter annihilation to hadronic modes.

\end{abstract}
\pacs{95.35.+d; 95.85.Ry; FERMILAB-PUB-09-033-A}
\maketitle

Recently, the PAMELA collaboration has published results which show an excess of cosmic ray positrons relative to electrons above approximately 10 GeV~\cite{PAMELA}, confirming previous indications from HEAT~\cite{Heat} and AMS-01~\cite{AMS}. Furthermore, the ATIC experiment has reported a surprising feature in the cosmic ray electron (plus positron) spectrum between approximately 300 and 800 GeV~\cite{ATIC}. These observations collectively indicate the presence of a bright source of very high energy electrons and positrons within a few kiloparsecs of the Solar System.

The leading astrophysical hypothesis for the origin of these particles is a nearby and relatively young pulsar (or pulsars), such as B0656+14 or Geminga~\cite{Dan,pulsars2,Stefano}.  A more exciting possibility, however, is that the signals reported by ATIC and PAMELA are the result of dark matter annihilations taking place in the halo of the Milky Way. In order for dark matter to generate these signals, however, care must be taken to avoid a number of potential problems. In particular, the spectrum of electrons and positrons predicted to be generated in the annihilations of most dark matter candidates is much too soft to fit the observations of PAMELA and ATIC~\cite{DanNeal}. Furthermore, if the annihilation rate throughout the halo of the Milky Way is normalized to account for the PAMELA and ATIC signals, most dark matter candidates will also generate an unacceptably large flux of cosmic ray antiprotons~\cite{antiprotons,antiprotons2}. Possible solutions to these problems include WIMPs which annihilate mostly to charged leptons~\cite{DanNeal,lepsom,leptons} or WIMPs which are distributed preferentially in the local neighborhood of the galaxy~\cite{clump}.

%Within the context of the particle physics model we discuss here, a dark matter candidate is predicted which simulataneously evades each of these potential problems.

In this letter, we consider Kaluza-Klein dark matter within the context of a model with one extra spatial dimension.  In this scenario, the extra dimension is universal, meaning that all of the Standard Model fields are free to propagate through it~\cite{Appelquist:2000nn}. Standard Model fields with momentum in the direction of the extra dimension appear as heavy particles known as Kaluza-Klein (KK) states. In particular, we assume that the extra dimension is compactified with an approximate size of $R \sim {\rm TeV}^{-1}$, leading to a complete KK copy of the Standard Model with masses at or around the TeV scale.

Although KK-number conservation is broken in phenomenologically viable scenarios, KK-parity can naturally be conserved in this model, leading to the stability of the lightest KK particle (LKP). An attractive choice for the LKP is the first KK excitation of the hypercharge boson, $B^{(1)}$. This state is stable, colorless, electrically neutral, and constitutes a viable candidate for dark matter~\cite{Servant:2002aq,Cheng:2002ej} (for a review, see Ref.~\cite{Hooper:2007qk}). $B^{(1)}$ pairs annihilate to (zero mode) fermion pairs through the $t$-channel exchange of a KK-fermion with a total cross section approximately given by $\sigma v \approx 1.7\times 10^{-26}\, {\rm cm}^3/{\rm s} \times  (1 \,{\rm TeV}/m_{B^{(1)}})^2$. If all other first level KK states are neglected, this cross section leads to a thermal relic abundance of KK dark matter equal to the measured density of dark matter for the choice of $m_{B^{(1)}} \approx 800$ GeV. If KK leptons and other KK states are included in the calculation, LKPs with masses in the range of approximately 600 to 900 GeV can easily be produced with the desired abundance~\cite{Servant:2002aq,coann}.

The $B^{(1)}-f^{(1)}-f^{(0)}$ couplings which go into the determination of the LKP's annihilation cross section each scale with the fermion's hypercharge, and thus annihilations to charged leptons are preferred over other possible final states. Furthermore, unlike neutralinos (or other Majorana fermions) the annihilation of KK dark matter to light fermions is not helicity suppressed. For a largely degenerate first level KK spectrum, LKPs annihilate to $e^+ e^-$, $\mu^+ \mu^-$ and $\tau^+ \tau^-$ 20\% of the time each. Of the remaining annihiations, most (approximately 35\%) produce quark pairs. The remaining fraction produce neutrinos and higgs bosons. As we will show, the annihilation modes of KK dark matter lead to very distinctive features in the cosmic ray positron fraction~\cite{kribs} and cosmic ray electron spectrum~\cite{baltz}.  Similar signatures are expected from other dark matter candidates which annihilate mostly to charged leptons.

To calculate the spectrum of cosmic ray electrons and positrons produced in the annihilations of KK dark matter, we solve the propagation equation~\cite{prop,BaltzEdsjo}:
\begin{eqnarray}
\frac{\partial}{\partial t}\frac{dN_{e}}{dE_{e}} &=& \vec{\bigtriangledown} \cdot \bigg[K(E_{e})  \vec{\bigtriangledown} \frac{dN_{e}}{dE_{e}} \bigg]
+ \frac{\partial}{\partial E_{e}} \bigg[b(E_{e})\frac{dN_{e}}{dE_{e}}  \bigg] \nonumber \\ &+& Q(E_{e},\vec{x}),
\label{dif}
\end{eqnarray}
where $dN_{e}/dE_{e}$ is the number density of electrons/positrons per unit energy, $K(E_{e})$ is the diffusion coefficient, and $b(E_{e}) \approx 10^{-16} \, ({E_e} / 1 \, \rm{GeV})^2 \,\, \rm{s}^{-1}$ is the electron/positron energy loss rate. As we expect to be in the steady state limit, we set the left hand side of Eq.~\ref{dif} to zero. The source term, $Q(E_e, \vec{x})$, reflects both the distribution of dark matter in the Galaxy, and the mass, annihilation rate, and dominant annihilation channels of the dark matter. We also must select boundary conditions for our diffusion zone.  In particular, we consider a cylinder of half-thickness $L$, beyond which cosmic rays are able to escape the Galactic Magnetic Field. For this calculation, we have used the package DarkSUSY~\cite{darksusy}.

\begin{figure}
\resizebox{8.5cm}{!}{\includegraphics{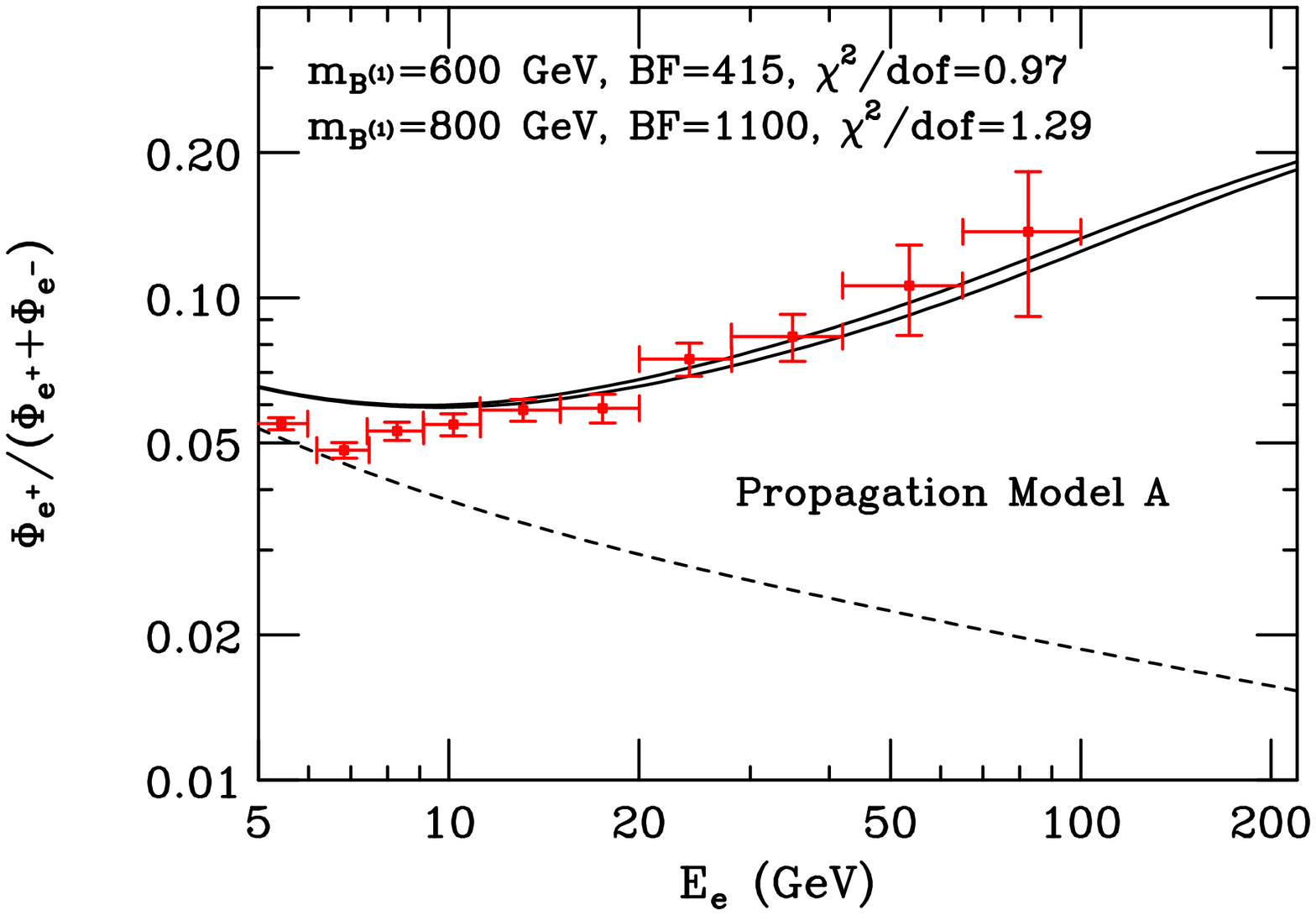}} \\
\resizebox{8.5cm}{!}{\includegraphics{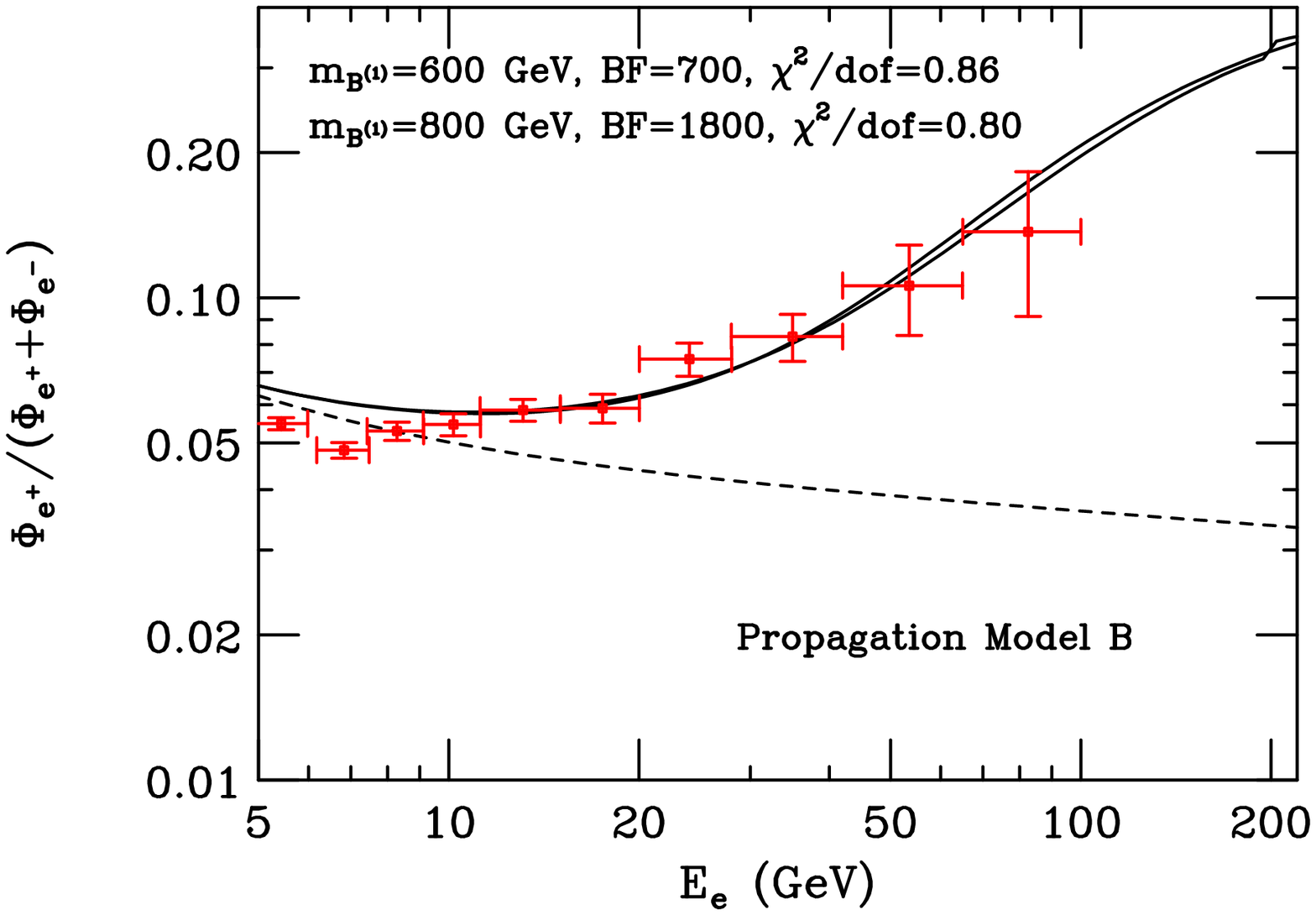}}
\caption{The positron fraction as a function of energy including contributions from Kaluza-Klein dark matter annihilations, compared to the measurements of the PAMELA experiment~\cite{PAMELA}. We show results for dark matter masses of 600 GeV and 800 GeV, and for two propagation models (see text for more details). In each frame, the dashed line denotes the positron fraction with no contribution from dark matter (secondary positron production only).}
\label{frac}
%\end{center}
\end{figure}

We consider two sets of diffusion parameters in our study which provide good fits to the measured B/C, sub-Fe/Fe, and Be$^{10}$/Be$^{9}$ ratios in the cosmic ray spectrum. We will refer to these two parameter sets as propagation models A and B:
\begin{eqnarray}
\mbox{A: $K(E_{e}) =5.3\times10^{28}\, (E_{e} / 4 \, \rm{GeV})^{0.43} \,{\rm cm^2/s}$, L=4\, {\rm kpc}  } \nonumber \\ 
\mbox{B: $K(E_{e}) =1.4\times10^{28}\, (E_{e} / 4 \, \rm{GeV})^{0.43} \,{\rm cm^2/s}$, L=1\, {\rm kpc}  }. \nonumber   
\end{eqnarray}

For the source term, we adopt a dark matter distribution which follows the Navarro-Frenk-White (NFW) halo profile~\cite{nfw}. We allow the normalization of the dark matter annihilation rate to be adjusted freely to fit to the observed PAMELA and ATIC features.

\begin{figure}
\resizebox{8.5cm}{!}{\includegraphics{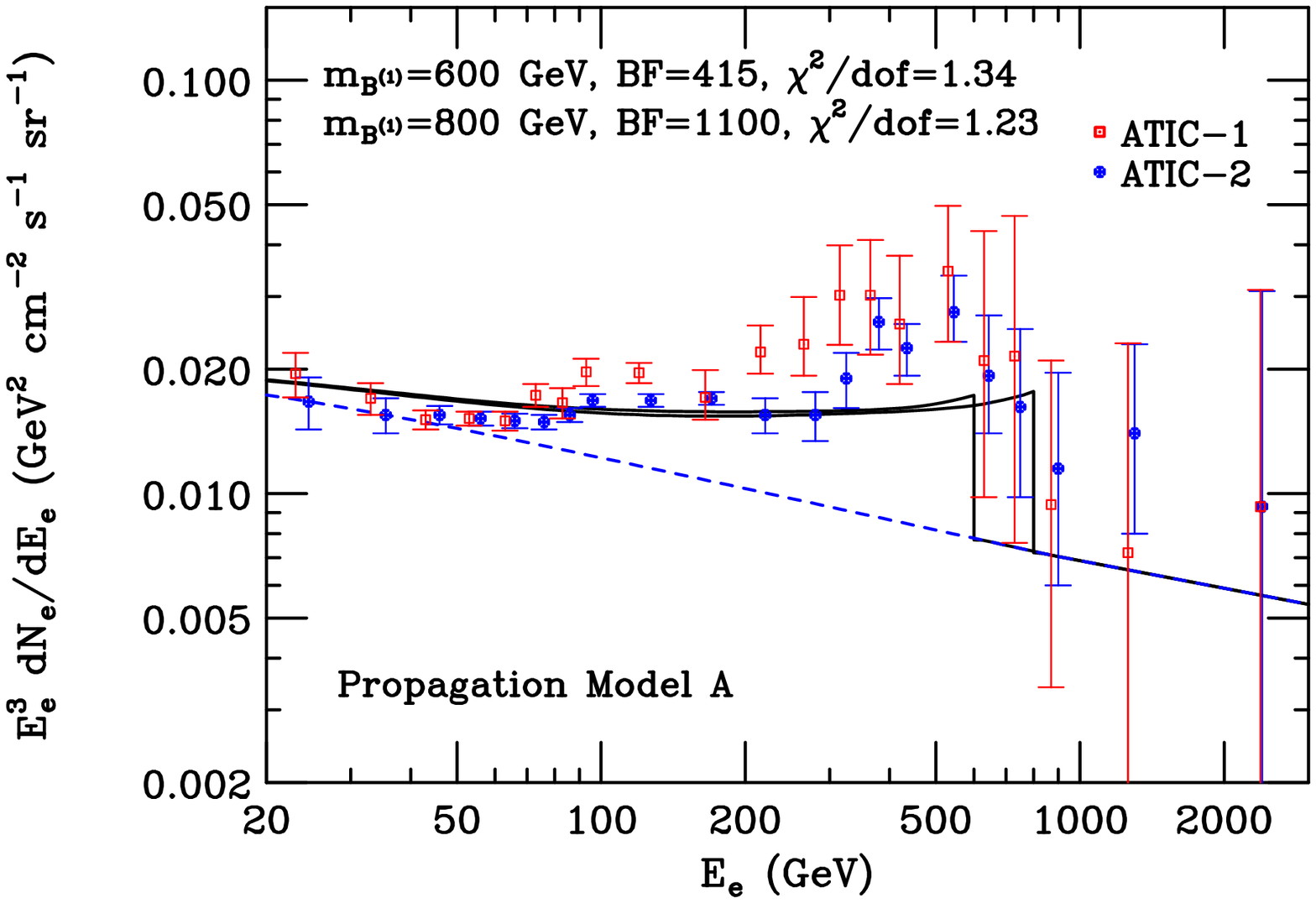}} \\
\resizebox{8.5cm}{!}{\includegraphics{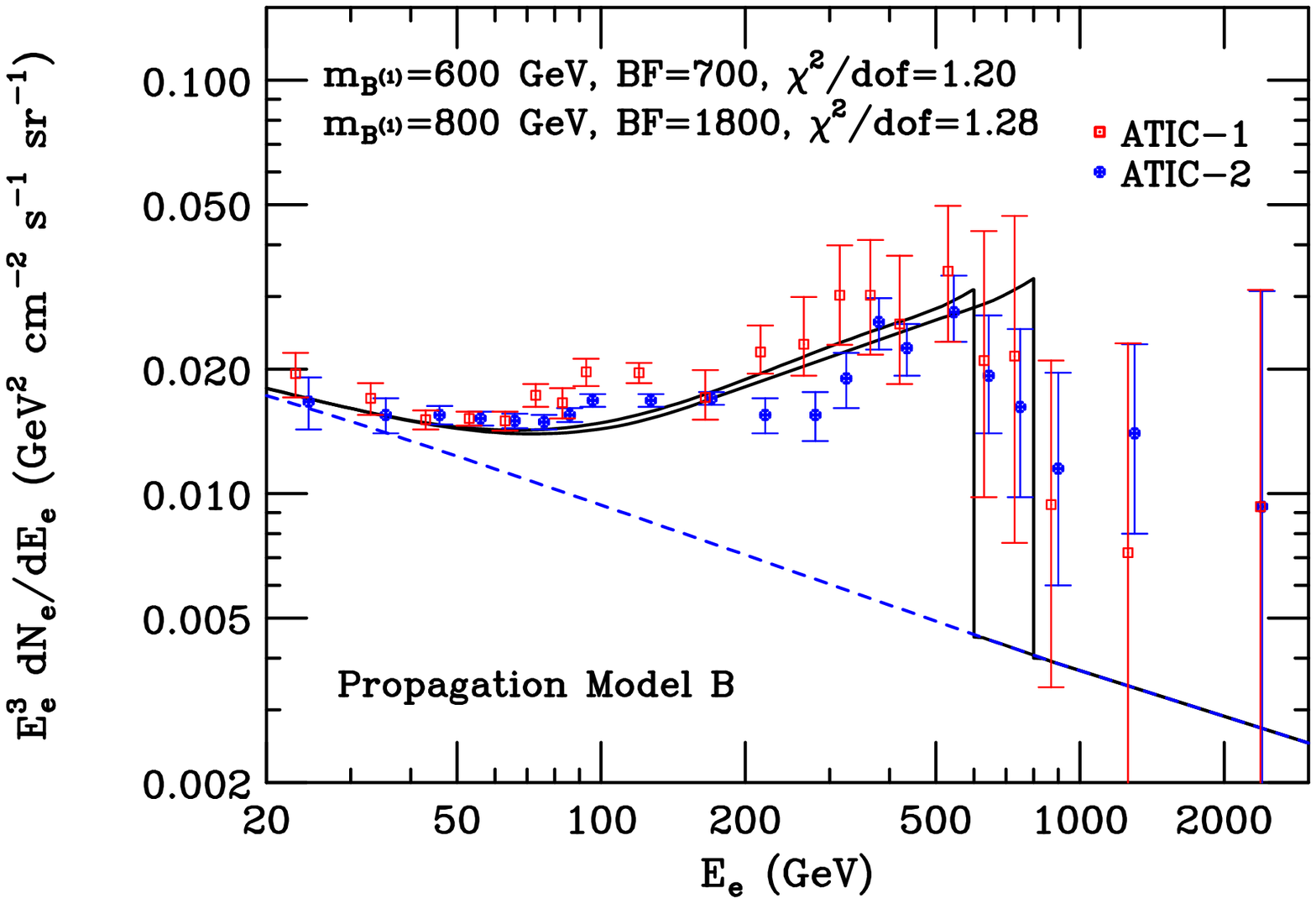}}
\caption{The electron plus positron spectrum including contributions from Kaluza-Klein dark matter, compared to the measurements of ATIC~\cite{ATIC}. We show results for dark matter masses of 600 GeV and 800 GeV, and for two propagation models (see text for more details). In each frame, the dashed line denotes the spectrum with no contribution from dark matter.}
\label{spec}
%\end{center}
\end{figure}

In Fig.~\ref{frac}, the positron fraction including the contribution from annihilating KK dark matter is compared to the measurements of PAMELA. In computing these results, we have adopted the cosmic ray electron and positron backgrounds as described in Ref.~\cite{MS,BaltzEdsjo}, but allow the spectral slope of the primary electron spectrum to vary within the current measurement errors. In particular, we adopt background electron spectral slopes of $dN_e/dE_e \propto E_e^{-3.25}$ and  $dN_e/dE_e \propto E_e^{-3.4}$ for propagation models A and B, respectively. We also allowed the normalization of the dark matter contribution ({\it ie.} the annihilation rate) to take on the value that best matches the data. In each case, we find excellent fits to the observations, although very high annihilation rates are required. We quantify this by introducing a boost factor (BF) which denotes the enhancement of the annihilation rate relative to a smooth halo with a local dark matter density of 0.3 GeV/cm$^3$. We find that boost factors in the range of approximately 400 to 2000 are required to produce the positron fraction observed by PAMELA. Although such values are considerably larger than those generally predicted based on the results of N-body simulations~\cite{vialactea}, such simulations do not have the resolution to study the small scale structure of the galactic halo distribution and rely on extrapolations to estimate annihilation boost factors. For this reason, we do not dismiss the possibility that small scale inhomogeneities in the dark matter distribution could significantly boost the annihilation rate.

In Fig.~\ref{spec}, we show the cosmic ray electron plus positron spectrum, with contributions from the same dark matter model and parameters shown in Fig.~\ref{frac}, compared to the measurements of ATIC~\cite{ATIC}. Here, we have attempted to account for possible systematic errors (which are not quantified in Ref.~\cite{ATIC}) by including 5\% systematic errors (in addition to the statistical errors shown) in the calculation of the $\chi^2$, and by allowing for an overall shift in the normalization/exposure of up to 25\%. Although very limited information is available regarding the systematic errors involved in these observations, we feel that we have made reasonable estimates of these quantities.

Based on the result shown in Figs.~\ref{frac} and~\ref{spec}, we conclude that KK dark matter is capable of providing a good fit to the combined observations of PAMELA and ATIC. In particular, we find acceptable values for the $\chi^2$ per degree of freedom for each mass and propagation model we have considered (although propagation model B provides somewhat better fits than model A). We would like to emphasize that the mass of the dark matter particle in this model is quite constrained by relic abundance considerations, and could not easily have been very different from the values we have considered here. In particular, to obtain a thermal relic abundance of KK dark matter, masses in the approximate range of 600 to 900 GeV are generally required~\cite{Servant:2002aq,coann}. Furthermore, KK masses lighter than about 300-400 GeV are inconsistent with electroweak precision measurements~\cite{lep}.

\begin{figure}
\resizebox{8.5cm}{!}{\includegraphics{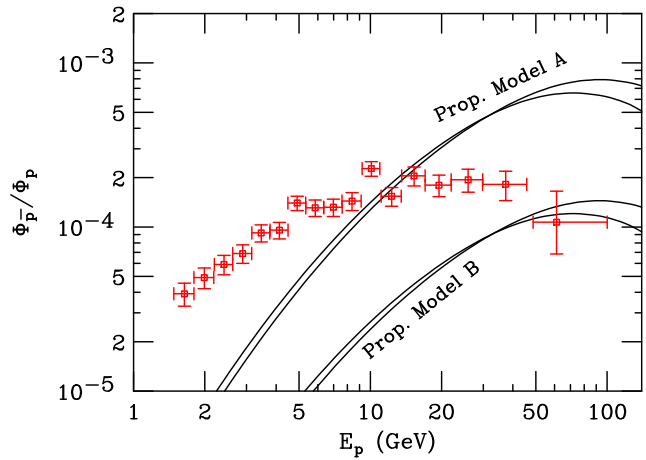}}
\caption{The contribution to the cosmic ray antiproton-to-proton ratio from annihilating Kaluza-Klein dark matter as normalized to the PAMELA and ATIC signals (and as in Figs.~\ref{frac} and~\ref{spec}). The upper two curves denote the results using propagation model A, and clearly exceed the antiproton content measured by PAMELA~\cite{pamelaantiproton}. The lower two lines denote the results for propagation model B and predict considerably fewer antiprotons, safely evading this constraint. Results are shown for dark matter masses of 600 and 800 GeV.}
\label{antiproton}
%\end{center}
\end{figure}

Before we can conclude that KK dark matter in this model is successful in producing the signals of PAMELA and ATIC, we must consider constraints from measurements of cosmic ray antiprotons, gamma rays, and synchrotron emission. In particular, dark matter annihilations which produce mostly quarks or gauge bosons are typically expected to produce more cosmic ray antiprotons than are observed by PAMELA~\cite{pamelaantiproton} if the overall annihilation rate is normalized to generate the PAMELA and ATIC positron/electron excesses. This problem is somewhat mitigated in the case of KK dark matter, however, as most of the annihilations proceed to charged leptons (which efficiently generate electrons and positrons, while not contributing to the antiproton flux). In Fig.~\ref{antiproton}, we show the contribution to the cosmic ray antiproton-to-proton ratio from KK dark matter, for the parameters (annihilation rates, masses, and propagation models) used in Figs.~\ref{frac} and~\ref{spec}. The upper two lines in this figure correspond to the results found using propagation model A, and clearly exceed the ratio measured by PAMELA~\cite{pamelaantiproton}. The lower two lines, in contrast, denote the results using propagation model B and are consistent with PAMELA's antiproton measurement. We thus conclude that in order for KK dark matter to produce the PAMELA and ATIC signals without overproducing antiprotons, a propagation model with a rather narrow diffusion zone ($L\sim 1$\,kpc) must be adopted. We remind the reader that such a propagation model is completely consistent with all current cosmic ray data.

Other constraints on dark matter annihilations in the Milky Way include those obtained by gamma ray, radio, and microwave observations of the Galactic Center region. If an NFW-like halo profile is adopted and the annihilation rate throughout the halo is enhanced by a common boost factor (normalized to the PAMELA and ATIC signals), these constraints are likely to be exceeded~\cite{CirelliStrumiaPho,syn}. As baryonic effects are anticipated to modify the dark matter distribution within the inner kiloparsecs (or parsecs) of the galaxy, however, it is difficult to arrive at any strong conclusions. In particular, the boost factors which enhances the dark matter annihilation rate could potentially vary with location, and could be considerably lower in the inner kiloparsecs of the Milky Way than in the vicinity of the Solar System. For example, if large boost factors exist, it would appear to imply the presence of a very large number of small sub-halos within the Milky Way. In the regions of the galaxy with the most stars ({\it ie.} the inner galaxy), tidal interactions are the most likely to destroy a large fraction of the sub-halos, leading to smaller boost factors relative to those further the Galactic Center~\cite{tidal}.

Instead of limiting our discussion to the specific KK dark matter candidate described here, we could retain much of the same phenomenology while considering a generic class of dark matter candidates which annihilate through couplings to hypercharge without chirality suppression and, thus, efficiently produce electrons and other charged leptons. In this more general context, we can imagine ways in which the need for a large boost factors could be relaxed.  In particular, dark matter particles with considerably larger cross sections than those implied by naive thermal abundance arguments are possible. For example, the dark matter particles may have been produced through a non-thermal mechanism in the early universe. Alternatively, the annihilation cross section could be strongly enhanced at low velocities due to non-perturbative processes~\cite{sommerfeld}, leading to a very high annihilation rate in the galactic halo. The results found for the case of KK dark matter in this letter can be applied to this broader classes of dark matter candidates, enabling the positron fraction and electron spectrum reported by PAMELA and ATIC to be generated without overproducing cosmic ray antiprotons, and potentially without the need of large boost factors.

In conclusion, we have shown here that Kaluza-Klein dark matter in a model with one universal extra dimension is capable of generating a spectrum of cosmic ray positrons and electrons consistent with the recent observations of PAMELA and ATIC. We have also shown that this can be accomplished without exceeding the measured quantity of cosmic ray antiprotons.

It is interesting to note that the mass of the Kaluza-Klein dark matter particle required to generate the observed dark matter abundance in this model is very similar to the value required to produce the feature in the cosmic ray electron spectrum reported by ATIC. Despite the overall lack of freedom in this model (the approximate mass and dominant annihilation channels are determined by the model), the PAMELA and ATIC signals can each be naturally accommodated. The least attractive feature of this scenario is the high annihilation rate ({\it ie.} large boost factor) that is required to normalize the overall signal to the PAMELA and ATIC data.

%%%

%Simulations \cite{vialactea,other} {\bf refs?} suggest $B \lesssim 10$, though it should be noted that the simulations do not yet have the range of scale to be able to conclude with certainty.  Lastly, a low velocity enhancement through the Sommerfeld  enhancement \cite{Nojiri,CirelliStrumiaMDM,TODM} or the Breit-Wigner \cite{BW} resonances, could lead to a boost. 

%To conclude, we showed that even with large annihilation cross-sections,  $B \sigma_{ann} v \simeq 10^{-23} \mbox{ cm}^3/\mbox{s}$, a UED model of dark matter can produce the ATIC and PAMELA signals, and be consistent with constraints from anti-protons and synchrotron radiation from the WMAP haze.  In the case of anti-protons, the consistency results in a model with a squeezed diffusion zone, lowering the fluxes of anti-protons by a factor $\sim 20$.

\medskip
We would like to thank Joakim Edsjo for him help with DarkSUSY. This work has been supported by the US Department of Energy, including grant DE-FG02-95ER40896, and by NASA grant NAG5-10842.

\end{document}